\documentclass[twocolumn]{aastex62}
\usepackage{graphicx}
\usepackage{amsmath}
\usepackage{amssymb}
\usepackage{dsfont}
\usepackage{textcomp}
\submitjournal{ApJ}
\begin{document}
\title{Stable equatorial ice belts at high obliquity in a coupled atmosphere-ocean model}
\correspondingauthor{Cevahir Kilic}
\email{kilic@climate.unibe.ch}

\author[0000-0002-3412-4429]{Cevahir Kilic}
\affil{Climate and Environmental Physics, Physics Institute, University of Bern, Switzerland}
\affil{Center for Space and Habitability, University of Bern, Switzerland}

\author{Frank Lunkeit}
\affil{Meteorological Institute, University of Hamburg, Germany}

\author[0000-0003-0176-0602]{Christoph C. Raible}
\affil{Climate and Environmental Physics, Physics Institute, University of Bern, Switzerland}
\affil{Center for Space and Habitability, University of Bern, Switzerland}
\affil{Oeschger Centre for Climate Change Research, University of Bern, Switzerland}

\author[0000-0003-1245-2728]{Thomas F. Stocker}
\affil{Climate and Environmental Physics, Physics Institute, University of Bern, Switzerland}
\affil{Center for Space and Habitability, University of Bern, Switzerland}
\affil{Oeschger Centre for Climate Change Research, University of Bern, Switzerland}

\begin{abstract}
% max 250 words

Various climate states at high obliquity are realized for a range of
stellar irradiance using a dynamical atmosphere-ocean-sea ice climate
model in an aquaplanet configuration. Three stable climate states are
obtained that differ in the extent of the sea ice cover. For low
values of irradiance the model simulates a Cryoplanet which has a
perennial global sea ice cover. By increasing stellar irradiance,
transitions occur to an Uncapped Cryoplanet with a perennial
equatorial sea ice belt, and eventually to an Aquaplanet with no
ice. Using an emulator model we find that the Uncapped Cryoplanet is a
robust stable state for a range of irradiance and high obliquities and
contrast earlier results that high-obliquity climate states with an
equatorial ice belt may be unsustainable or unachievable. When the
meridional ocean heat flux is strengthened, the parameter range
permitting a stable Uncapped Cryoplanet decreases due to melting of
equatorial sea ice. Beyond a critical threshold of meridional ocean
heat flux, the perennial equatorial ice belt disappears. Therefore, a
vigorous ocean circulation may render it unstable. Our results suggest
that perennial equatorial ice cover is a viable climate state of a
high-obliquity exoplanet. However, due to multiple equilibria, this
state is only reached from more glaciated, but not from less glaciated
conditions.

\end{abstract}

%% Keywords should appear after the \end{abstract} command. 
%% See the online documentation for the full list of available subject
%% keywords and the rules for their use.

\section{Introduction}

Orbital configurations such as obliquity and stellar irradiance are
key determinants of the surface temperature of a planet
\citep[e.g.,][]{Pierrehumbert2010,Ferreira2014,Kaspi2015,Kilic2017b}. Different
combinations of these variables lead to a variety of states, such as a
completely ice covered Cryoplanet, an ice free Aquaplanet, or
combinations of these
\citep{Ferreira2014,Kilic2017b,Rose2017}. \citet{Ferreira2014}
explored the role of the ocean in controlling the surface temperature
under high obliquity and suggested that the climate state with a
perennial equatorial ice cover (Uncapped Cryoplanet) is unlikely to be
stable due to the meridional structure of the ocean heat
flux. Likewise, \citet{Rose2017} concluded based on an analytical
latitude-energy balance model that a state with an equatorial ice belt
may be stable but could not be reached in the presence of other
equilibria such as ice-free and snowball states. On the other hand,
\citet{Kilic2017b} using an atmospheric general circulation model
(AGCM) coupled to a slab ocean found a stable uncapped cryoplanet
state. This state was found stable for obliquities above
54\textdegree\ and a range of stellar irradiances. They pointed to
the importance of initial conditions: in their simulations an Uncapped
Cryoplanet was obtained only when the model was initialized in a
cryoplanet state before increasing stellar irradiance. In the entire
parameter region of the Uncapped Cryoplanet, there also exists the
Aquaplanet as a second stable state. This implies
that the state with an equatorial ice belt could not be reached from
an aquaplanet initial state. This dependence on initial conditions is
a characteristic of hysteresis behavior and the consequence of the
broad occurrence of multiple equilibria in the obliquity-irradiance
parameter space as shown by \citet{Kilic2017b}. However, the precise
shape of the hysteresis is strongly model and parameter dependent.

In this study we explore climate states at high obliquity using an
AGCM coupled to a fully dynamical ocean model. In particular, we
assess the role of the ocean heat transport for the uncapped
cryoplanet state. The analysis focuses on the meridional energy flux
of the atmosphere and the ocean which together determine the ice-free
poles and the ice covered equatorial band. The results obtained by the
dynamical atmosphere-ocean-sea ice model are complemented by a large
number of simulations with an emulator model that consists of a
dynamical atmosphere coupled to a slab ocean. The emulator, on which
the earlier study of \citet{Kilic2017b} was based, permits us here to
specifically explore the stability of the Uncapped Cryoplanet at
obliquities above 54\textdegree\ with respect to increasing meridional
ocean heat flux and to investigate whether such a state may be a likely
climate state of an exoplanet.

Earth may have been in a high obliquity state during its early history
\citep{Williams1975}, and this has evoked a number of studies which
employed a hierarchy of different AGCMs coupled to a slab ocean and a
thermodynamic sea ice model \citep[e.g.,][]{Jenkins2003}. Also, an
Earth partially or even fully covered by ice, the so-called Snowball
Earth, was likely a climate state for some time in the deep past
\citep[e.g.][]{Pierrehumbert2011}. While more recent studies have
suggested that a high-obliquity state of Earth may have been unlikely
\citep{Levrard2003}, exoplanet research has become interested in
high-obliquity configurations
\citep{Williams2003,Armstrong2014,Ferreira2014}. In particular,
planets without a sizeable moon lack an important stabilizer of the
planetary rotation axis and may experience large variations of
obliquity \citep{Laskar1993}. \citet{Abe2011} showed that for dry
(land) planets high obliquity can lead to snow accumulation at
low-latitudes, whereas studies with oceans concluded that a stable
state with equatorial sea ice is unlikely
\citep[e.g.,][]{Ferreira2014}, or unachievable
\citep{Rose2017}. \citet{Kilic2017b} used an atmospheric model coupled
to a slab ocean and found six distinct climate states, including the
Uncapped Cryoplanet, in the obliquity-stellar irradiance parameter
space. Because of the existence of multiple equilibria in their model,
some states are only reached when starting from specific initial
conditions: transitions to an uncapped cryoplanet state only occurred
when simulations started from a cryoplanet state and could so escape
detection in simulations that only start from a single initial state.

The paper is organized as follows. Section \ref{sec:Model} describes
the climate model and the experimental setup. Section
\ref{sec:Circulation} presents the atmospheric and ocean dynamics for
different climate states. In Section \ref{sec:stable_regimes} we
explore the parameter space for a stable Uncapped
Cryoplanet. Discussion and conclusion are presented in Section
\ref{sec:Conclusion}.

\section{Model and experimental design}\label{sec:Model}

The study is based on simulations performed with the Planet Simulator
(PlaSim) \citep{Lunkeit2011} in an aquaplanet configuration, i.e.,
without continents.  \citep{Kilic2017a,Kilic2017b}. It is a climate
model of intermediate complexity consisting of an AGCM, which can be
coupled to a dynamical 3-dimensional ocean (version OD) or to a slab
ocean (version OS). In OS the ocean component is a single water layer,
a slab of 50\,m thickness, and the horizontal transport of heat is
parameterized by eddy diffusion with a horizontal diffusivity
$K_h$. Both model configurations feature a thermodynamic sea ice
model.

The ocean component of OD is the Hamburg LSG model
\citep{Maier-Reimer1993,Prange2003} which integrates the momentum
equations forward in time, including all terms except the nonlinear
advection of momentum, by an implicit scheme thereby allowing for a
large time step. The model has 22 vertical levels of 50 to 1000\,m
spacing and a flat bottom at a depth of 5500\,m. The model is
formulated on a semi-staggered E-grid with $72\times76$ grid cells and
is in the aquaplanet configuration, i.e. there are no continents
\citep{Hertwig2015}. The atmospheric component of PlaSim in version OD
has a spectral resolution of T21 (approximately 5.6\textdegree) and 10
vertical levels.

A set of simulations with OD is carried out to determine the parameter
range in which an Uncapped Cryoplanet is sustained
(Fig.\ \ref{fig:OD_atm}). Simulations start from a cryoplanet state
with $\epsilon = 90$\textdegree\ and $\tilde{S} = 0.70$ ($S =
\tilde{S}\cdot S_0$ with $S_0 = 1361$\,W\,m$^{-2}$). Upon increasing
the stellar irradiance to $\tilde{S} = 0.94$ an Uncapped Cryoplanet is
obtained in OD. An Aquaplanet is realized from an Earth-like control
simulation ($\epsilon = 23.44$\textdegree, $\tilde{S} = 1$) by
increasing $\epsilon$ to 90\textdegree. These states are close to
equilibrium as indicated by nearly vanishing trends in global mean
ocean temperature ($<$10$^{-6}$\,K\,yr$^{-1}$).

In version OS, effects of ocean circulation cannot be studied, but
there is the possibility to vary the relative importance of the
meridional ocean heat flux by varying the horizontal diffusivity of
the slab ocean. The atmospheric component of PlaSim in version OS has
a spectral resolution of T42 (approximately 2.8\textdegree) and 10
vertical levels. We show that OS can be used as an emulator model of
the computationally more expensive version OD.

With the emulator we sample the parameter space of obliquity
$\epsilon$, stellar irradiance $\tilde{S}$, and horizontal diffusivity
$K_h$. All simulations start from the cryoplanet state and are run
into equilibrium. Obliquity is varied from 55\textdegree\ to
90\textdegree, stellar irradiance $\tilde{S}$ from 0.93 to 1.10, and
$K_h$ from 0 to $15\times 10^4$\,m$^2$\,s$^{-1}$. Simulations were
integrated for 100 up to 500 years to verify stability of the
state. In total 308 simulations were carried out.

\section{Atmospheric and oceanic circulation}\label{sec:Circulation}

\begin{figure*}[p]%[tttbbb]
    \centering {
      \includegraphics[width=1.\linewidth]{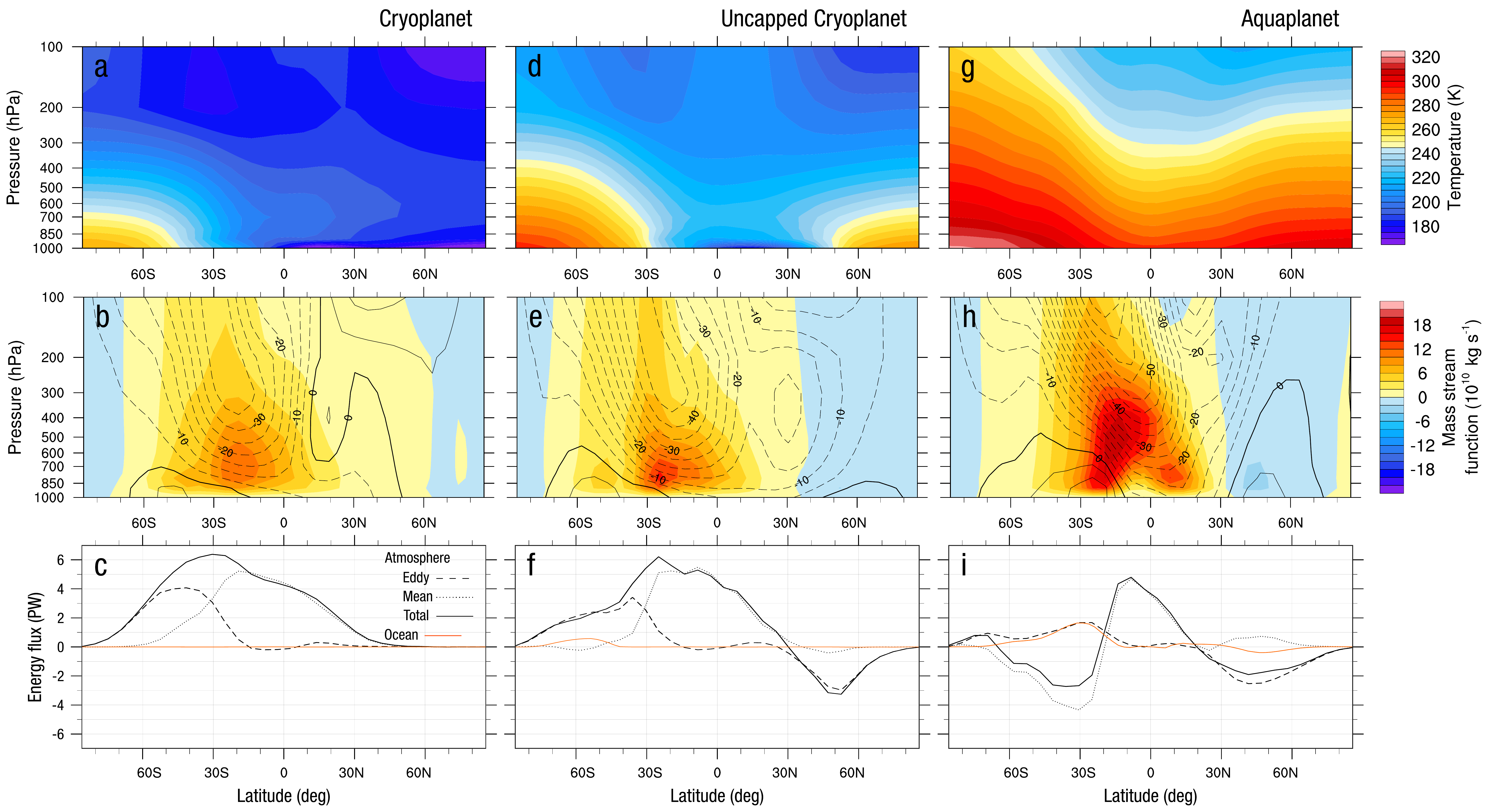} }
\caption{Stable climate states simulated using a coupled
  atmosphere-ocean-sea ice model (version OD) at high obliquity
  ($\epsilon = 90$\textdegree). Shown are ten-year monthly means for
  January. a-c: Cryoplanet (relative stellar irradiance $\tilde{S} =
  0.7$). d-f: Uncapped Cryoplanet ($\tilde{S} = 0.94$). g-i:
  Aquaplanet ($\tilde{S} = 1.0$). Shown are (a,d,g) the zonal-mean
  atmospheric temperature (color interval 5\,K), (b,e,h) mass stream
  function (color interval $2\times10^{10}$\,kg\,s$^{-1}$, positive
  for clockwise overturning) and zonal-mean zonal wind (contour
  interval 5\,m\,s$^{-1}$, positive for westerlies), and (c,f,i) the
  meridional atmospheric energy flux, consisting of the eddy and mean
  circulation contributions, and the meridional ocean heat
  flux. Positive values represent a northward flux.}\label{fig:OD_atm}

%\begin{figure*}[!ttt]
    \centering {
      \includegraphics[width=1.0\linewidth]{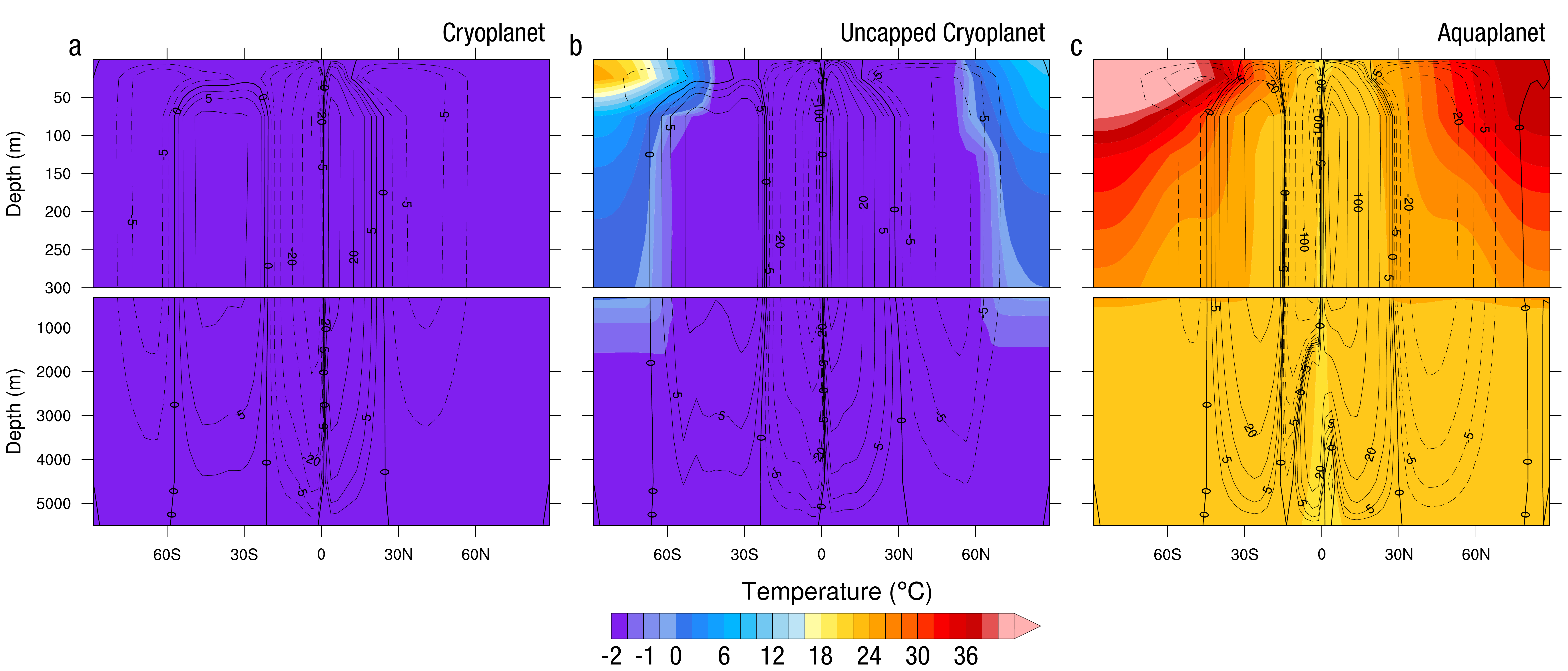} }
    \caption{Ocean component of the stable climate states at high
      obliquity ($\epsilon = 90$\textdegree) shown in
      Fig.\ \ref{fig:OD_atm}. a: Cryoplanet ($\tilde{S} = 0.7$). b:
      Uncapped Cryoplanet ($\tilde{S} = 0.94$). c: Aquaplanet
      ($\tilde{S} = 1.0$). Shown are the zonal-mean ocean
      temperature (color interval: 0.5\textcelsius\ for $ <
      0$\textcelsius\ and 2\textcelsius\ for $ \geq 0$\textcelsius)
      and the meridional overturning (contour interval are in Sv:
      $0,\pm2.5,\pm5,\pm10,\pm20,\pm50,\pm100,\pm150,\pm200$; positive
      for clockwise overturning).}\label{fig:OD_oce}
%\end{figure*}

\end{figure*}

At high obliquity $\epsilon$, the fully dynamical model OD exhibits
three different climate states depending on the strength of the
stellar irradiance. This is shown for $\epsilon =
90$\textdegree\ where for increasing values of $\tilde{S}$, the
Cryoplanet, Uncapped Cryoplanet, and the Aquaplanet are found
(Fig.\ \ref{fig:OD_atm}). This is consistent with, and confirms
earlier simulations using the simpler model version with only a slab
ocean \citep[OS,][]{Kilic2017b}. We note, however, that the Near
Cryoplanet, as state with seasonally open polar oceans, is not
obtained with the OD version. Nevertheless, this similar solution
structure at $\epsilon = 90$\textdegree\ indicates that OS is a
reasonable emulator of OD. This will be corroborated further
below.

The main characteristics of the climate states are shown in
Fig.\ \ref{fig:OD_atm} for January. The atmosphere of the Cryoplanet
is very cold due to the sea ice albedo reflecting large amounts of the
stellar energy flux back to space. Only the polar atmosphere facing
the star is significantly warmer than the remaining part of the planet
highlighting the strong seasonality of the cryoplanet climate. The
Uncapped Cryoplanet has an ice cover that is confined to the low
latitudes, and both polar areas are ice free. This is due to the
underlying open ocean that absorbs heat during the summer months which
keeps the ocean surface ice free in the winter. Therefore, both polar
atmospheres are warm throughout the year. In the aquaplanet state
surface temperatures are warm at all latitudes with a seasonal cycle
that is damped by the presence of the underlying ocean.

The zonal circulation in the atmosphere shows a strong
thermally-driven easterly jet in all three states in the summer
hemisphere (Fig.\ \ref{fig:OD_atm}).  Still, differences in structure,
strength and mechanisms are present. In the case of the Cryoplanet and
Uncapped Cryoplanet, the overturning circulation of the atmosphere is
driven by both eddy and mean atmospheric energy fluxes \cite[for
  further details see][]{Kilic2017b}. For the Aquaplanet, the
overturning circulation consists of two thermally direct driven cells
in the area of 60\textdegree S to 20\textdegree N. The thermally
indirect atmospheric circulation in the winter hemisphere of the
Aquaplanet is caused by the dominant northward eddy heat flux
(Fig.\ \ref{fig:OD_atm}i).

Apart from the strong easterly jet, the zonal circulation in the
atmosphere differs in structure and strength in the three
states. Meridionally, all three states feature a thermally direct
one-cell circulation in the summer hemisphere. In the case of the
Cryoplanet and Uncapped Cryoplanet the underlying mechanism driving
the atmospheric circulation is the same \cite[for further details
  see][]{Kilic2017b}.

Ocean circulation contributes very little to the meridional
heat flux when the planet is completely or partly ice covered
(Fig.\ \ref{fig:OD_atm}c,f). Conversely, there is a substantial ocean
heat flux in the summer hemisphere towards the equator in the case of
the Aquaplanet (Fig.\ \ref{fig:OD_atm}i).

%\begin{figure*}[!ttt]
%    \centering {
%      \includegraphics[width=1.0\linewidth]{kilic_2018_Fig2.pdf} }
%    \caption{Ocean component of the stable climate states at high
%      obliquity ($\epsilon = 90$\textdegree) shown in
%      Fig.\ \ref{fig:OD_atm}. a: Cryoplanet ($\tilde{S} = 0.7$). b:
%      Uncapped Cryoplanet ($\tilde{S} = 0.94$). c: Aquaplanet
%      ($\tilde{S} = 1.0$). Shown are the zonal-mean ocean
%      temperature (color interval: 0.5\textcelsius\ for $ <
%      0$\textcelsius\ and 2\textcelsius\ for $ \geq 0$\textcelsius)
%      and the meridional overturning (contour interval are in Sv:
%      $0,\pm2.5,\pm5,\pm10,\pm20,\pm50,\pm100,\pm150,\pm200$; positive
%      for clockwise overturning).}\label{fig:OD_oce}
%\end{figure*}
%
Figure \ref{fig:OD_oce} shows the zonal-mean temperature and
meridional overturning in the ocean for the simulations given in
Fig.\ \ref{fig:OD_atm}. Below 100\,m depth, the ocean temperature is
nearly symmetric about the equator despite the strong seasonal
variability of the atmosphere at $\epsilon =
90$\textdegree\ (Fig.\ \ref{fig:OD_atm}). This is due to the large
heat capacity of the ocean and the associated thermal adjustment time
on the order of 10$^3$~years.

In the case of the cryoplanet and the uncapped cryoplanet states, the
ocean temperature below the sea ice cover is close to the freezing
point and features very small meridional and vertical gradients.
Therefore, the meridional overturning is mainly caused by the Ekman
transport driven at the surface by a combination of wind stress and
its transfer through sea ice (Fig.\ \ref{fig:OD_oce}a,b). Overturning
strengthens towards the equator because Ekman transport scales with
the inverse of the Coriolis parameter. The ocean temperature of the
Uncapped Cryoplanet reflects the meridional structure of surface
temperature in the atmosphere with the warmest water in the ice-free
polar areas (Fig.\ \ref{fig:OD_oce}b). The polar thermocline reaches
to a depth of about 400\,m in winter and less than 100\,m in
summer. Under the sea ice, vertical temperature gradients are very
small, which explains the very small meridional ocean heat flux
despite a substantial meridional overturning circulation
(Fig.\ \ref{fig:OD_atm}c,f).

In the case of the aquaplanet state, the meridional overturning in the
ocean is additionally driven by the density gradients as is evident
from the large vertical and meridional temperature (and salinity)
gradients (Fig.\ \ref{fig:OD_oce}c). This leads to a substantial
equatorward meridional heat flux in the ocean in the summer hemisphere
(Fig.\ \ref{fig:OD_atm}i).

\newpage

\section{Stable regimes and Hysteresis}\label{sec:stable_regimes}

\begin{figure*}[tttbbb]
    \centering {
      \includegraphics[width=1.\linewidth]{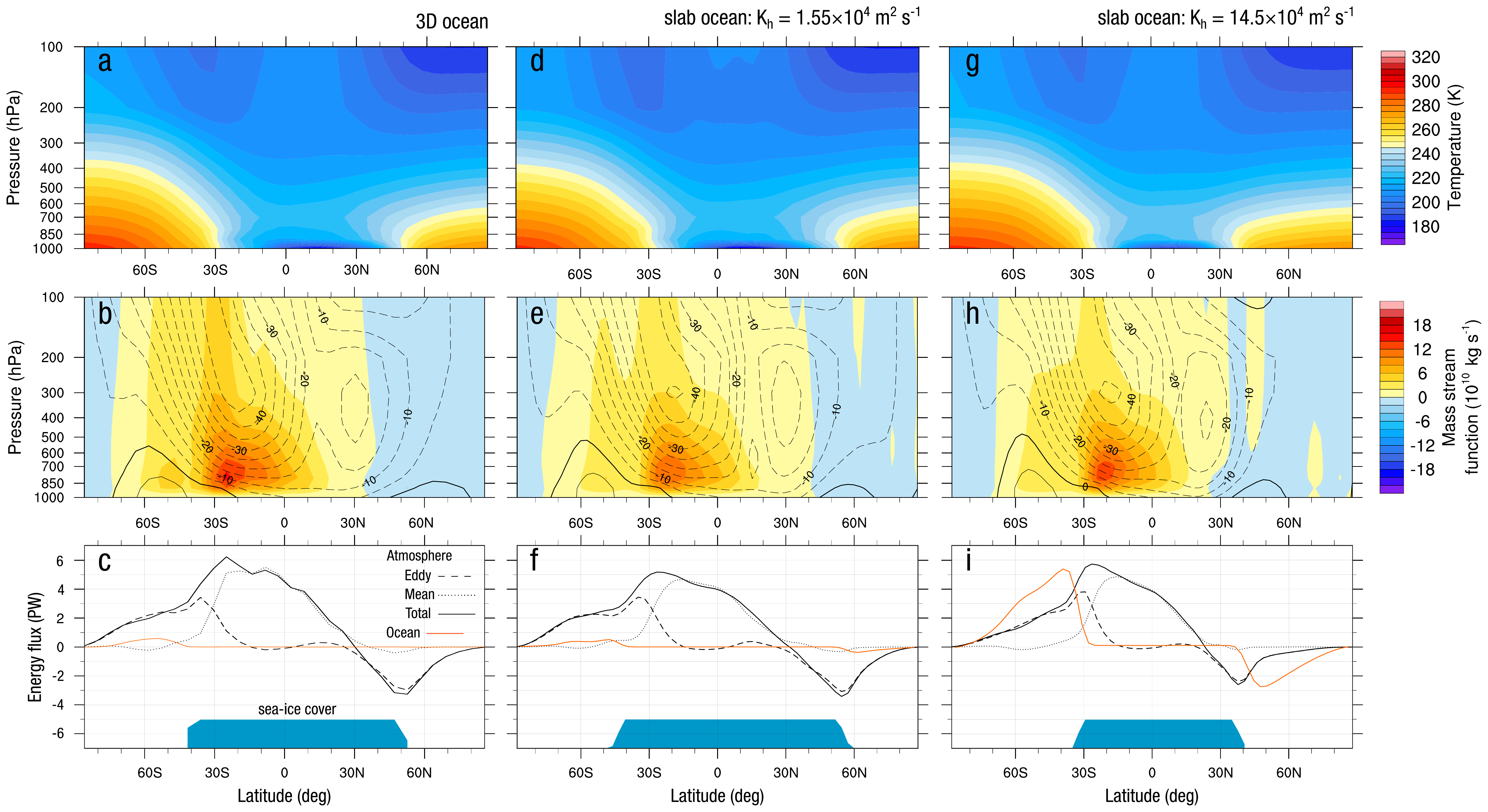} }
\caption{Same as Fig.\ \ref{fig:OD_atm} but for three uncapped
  cryoplanet states showing ten-year monthly means for January with
  $\epsilon = 90$\textdegree\ and $\tilde{S} = 0.94$: (a-c) OD
  configuration as in Fig.\ \ref{fig:OD_atm}. (d-f) OS configuration
  $K_h = 1.55\cdot10^4$\,m$^2$\,s$^{-1}$. (g-i) OS configuration with
  $K_h = 14.5\cdot10^4$\,m$^2$\,s$^{-1}$. In the bottom panels, the
  blue areas indicate the sea ice cover. Note that the OS
  configuration (d-f) is a good emulator of the dynamical model
  configuration OD. In OS the meridional ocean heat flux can be
  increased by increasing $K_h$ (i).}\label{fig:UC_atm}
\end{figure*}

An earlier study suggested that a dynamical ocean circulation would
not permit stable climate states with a perennial equatorial ice
cover, i.e., an uncapped cryoplanet state, under high obliquity
conditions \citep{Ferreira2014}. They started their simulations from
an aquaplanet configuration. Not inconsistent with this finding, our
earlier simulations using OS \citep{Kilic2017b} showed that an
uncapped cryoplanet state could only be reached from a cryoplanet but
not from an aquaplanet configuration. Here we find that such a state
may also in a fully dynamical atmosphere-ocean model (OD). In order to
investigate the sensitivity of the uncapped cryoplanet state to the
strength of the meridional ocean heat transport we now use OS as an
emulator. In OS the relative importance of the ocean heat transport
can be varied by the horizontal diffusivity $K_h$ of the slab
ocean. This is illustrated in Fig.\ \ref{fig:UC_atm} for the uncapped
cryoplanet state at $\epsilon$ = 90\textdegree\ and
$\tilde{S}=0.94$. With a horizontal diffusivity of $K_h =
1.55\cdot10^4$\,m$^2$\,s$^{-1}$ the emulator OS reproduces the
seasonal distribution of atmospheric temperature, the zonal and
meridional atmospheric circulations, as well as the meridional heat
fluxes in the ocean and the atmosphere, including the eddy and mean
components. Also the equatorial sea ice cover simulated by the
emulator is in quantitative agreement with the results from the
dynamical atmosphere-ocean model (Fig.\ \ref{fig:UC_atm}a-c
v.s. d-f).

In the emulator we can strengthen the meridional ocean heat flux by
simply increasing $K_h$. This simulates a case in which the effect of
the distribution of heat via the ocean circulation becomes more
important.  If the horizontal diffusivity is increased by almost an
order of magnitude to $K_h = 14.5\cdot10^4$\,m$^2$\,s$^{-1}$, the peak
meridional ocean heat flux towards the ice edges reaches a magnitude
similar to that in the atmosphere (Fig.\ \ref{fig:UC_atm}i). In spite
of this large supply of heat to the ice edge, the Uncapped Cryoplanet
remains a steady state. However, a slight further increase of the slab
ocean diffusivity $K_h$, and in consequence of the meridional ocean
heat flux, causes the crossing of a tipping point which triggers the
irreversible melting of the equatorial sea ice with subsequent
transition to the aquaplanet state. It is the ocean heat flux towards
the ice edge that eventually destroys the equatorial ice belt.

\begin{figure*}[ttt]
    \centering {
      \includegraphics[width=1.0\linewidth]{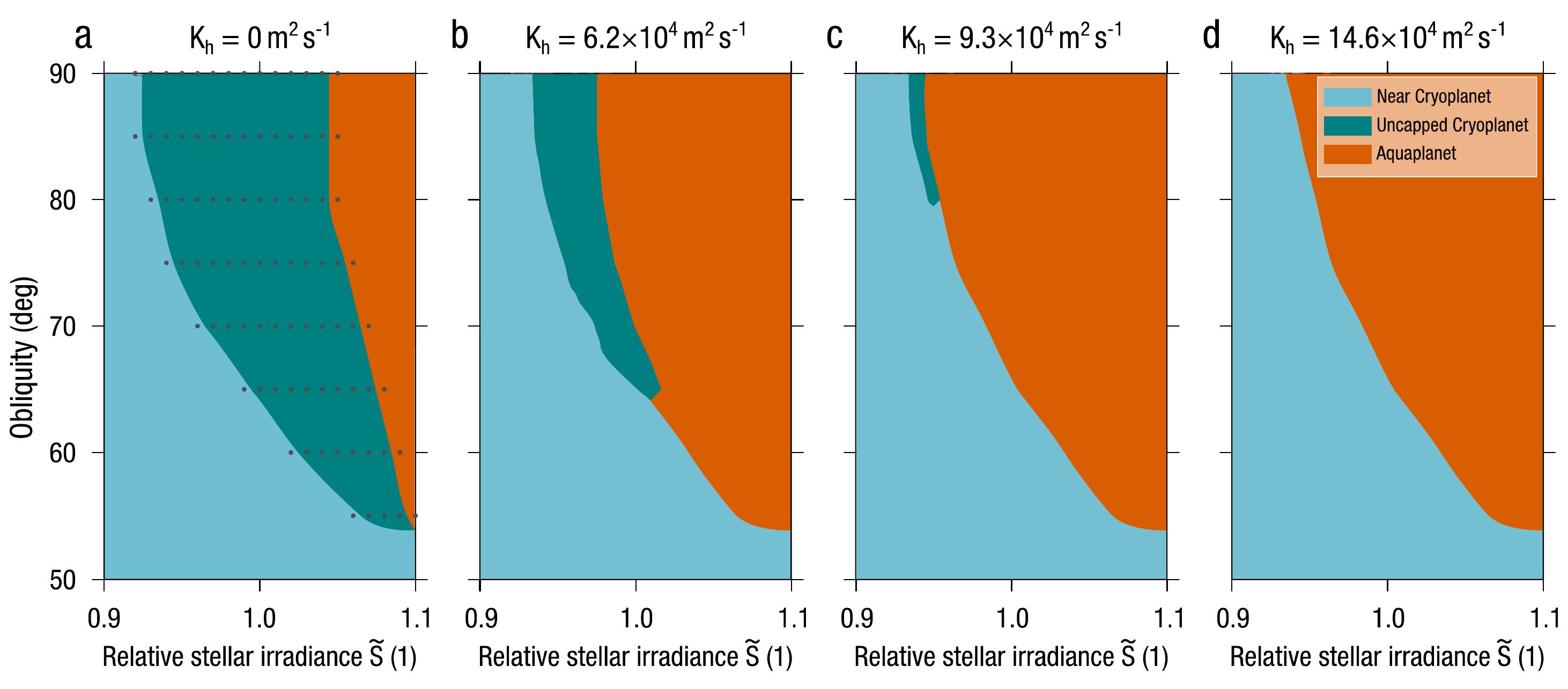} }
\caption{Regions of three distinct stable climate states in
  obliquity-irradiance space using a dynamical atmosphere coupled to
  slab ocean and sea ice model (OS configuration). All simulations are
  started from a cryoplanet initial state. The dots in panel (a)
  indicate the sampled locations in obliquity-irradiance
  space. Increasing diffusivities $K_h$ of the slab ocean were used
  for these simulations to determine the effect on the location of the
  region boundaries of the uncapped cryoplanet state. The area where
  the Uncapped Cryoplanet is stable is largest in the configuration of
  a swamp ocean ($K_h=0$\,m$^2$\,s$^{-1}$, panel a) and decreases with
  increasing $K_h$. If $K_h$ exceeds a critical value ($K_h \geq 14.6
  \cdot 10^4$\,m$^2$\,s$^{-1}$), only the Aquaplanet and the Near
  Cryoplanet remain as stable states (panel d).}\label{fig:UC_range}
\end{figure*}

With the OS emulator we now explore the region of stability for the
uncapped cryoplanet state in the
$\epsilon$-$\tilde{S}$-$K_h$-parameter space. In
Fig.\ \ref{fig:UC_range} we show the change of the steady state
regions with increasing $K_h$. Sampling of the parameter space using
the emulator OS permits us to locate the state boundaries. In the
range from $K_h=0$\,m$^2$\,s$^{-1}$ (no meridional heat flux in the
ocean) to $K_h=9.6\cdot10^4$\,m$^2$\,s$^{-1}$ we find three different
equilibrium states. Generally, for high $\tilde{S}$, no ice is
sustained and an Aquaplanet results; for lower $\tilde{S}$ a Near
Cryoplanet is simulated which has a seasonally open polar area. In
between, the Uncapped Cryoplanet is found which has perennially open
polar areas and an equatorial ice belt. This is a stable state that is
robust against rather large changes of horizontal diffusivity $K_h$
around the standard value. The choice of the standard $K_h =
1.55\cdot10^4$\,m$^2$\,s$^{-1}$ for the OS model is based on
present-day conditions, i.e.\ $\tilde{S}=1$ and $\epsilon =
23.4$\textdegree, and agreement with the large-scale atmospheric
properties and dynamics \citep{Kilic2017b}. By increasing $K_h$ by
almost an order of magnitude this state remains stable
(Fig.\ \ref{fig:UC_range}c), although the extent of the region of
existence reduces significantly.

For larger $K_h$, and hence stronger ocean heat transport, the
uncapped cryoplanet region is confined to a narrower range of
$\tilde{S}$ and higher obliquity. For $K_h =
14.5\cdot10^4$\,m$^2$\,s$^{-1}$ the uncapped cryoplanet remains a
stable state at just $\epsilon$=90\textdegree\ and $\tilde{S}=0.94$
(Fig.\ \ref{fig:UC_atm}g-i). For even larger $K_h$ this state is no
longer realized, as test simulations at $K_h =
20\cdot10^4$\,m$^2$\,s$^{-1}$ have revealed. This defines a critical
threshold of $K_h$ beyond which the meridional ocean heat flux is too
large to sustain equatorial ice. Our simulations suggest that this
threshold is at about $14.6 \cdot 10^4$\,m$^2$\,s$^{-1}$ for the
present model configuration.

For the swamp ocean that only acts as a heat storage ($K_h =
0\,$m$^2$\,s$^{-1}$), the stable regime for the uncapped cryoplanet
state reaches its largest extent. This highlights the sensitivity of
the equilibrium states on the repartitioning of meridional energy
fluxes. With increasing dominance of the ocean transport, the uncapped
cryoplanet state is sustained in only a small area of the parameter
space before it loses stability. Therefore, it depends on the details
of the ocean circulation how many different equilibrium states of
planetary climate can be realized.

\begin{figure*}[ttt]
    \centering {
      \includegraphics[width=1.0\linewidth]{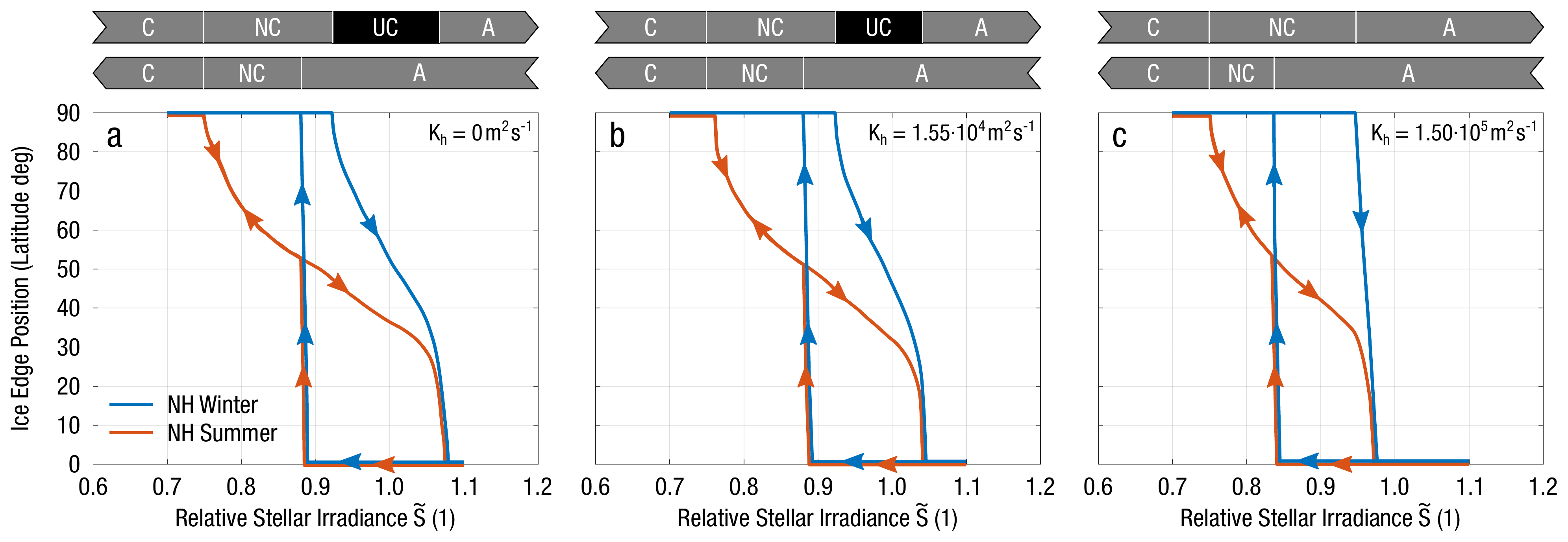} }
    \caption{ Hysteresis loops of the northern winter and summer ice
      edge positions in the Northern Hemisphere as the relative
      stellar irradiance $\tilde{S}$ is increased or
      decreased. Simulations are carried out with the OS version at
      $\epsilon=90$\textdegree\ for three values of $K_h$ as shown in
      panels (a), (b) and (c). The upper horizontal gray bars above
      the panels indicate the equilibrium states that are visited
      successively when increasing $\tilde{S}$: C Cryoplanet (complete
      ice cover), NC Near Cryoplanet (seasonally open polar oceans),
      UC Uncapped Cryoplanet (perennially open polar oceans), and A
      Aquaplanet (ice free). The sequence of equilibria is reversed
      when decreasing $\tilde{S}$, as shown in the lower horizontal
      gray bars, except that no UC is simulated because of a tipping
      point in the winter ice edge position.  The shape of the
      hysteresis loops for the ice edge positions, and hence the
      sequence of equilibrium states, depends on $K_h$; for large
      $K_h$ (panel c), no UC is simulated in either direction of
      $\tilde{S}$ change. For decreasing $\tilde{S}$ no UC is simulated
      independent of the value of $K_h$. }\label{fig:hysteresis}
\end{figure*}

In order to illustrate how the initial condition influences the stable
equilibrium, we select obliquity $\epsilon=90$\textdegree\ and use the
OS version for three cases of $K_h$. By slowly increasing $\tilde{S}$
by $10^{-4}$\,yr$^{-1}$, equilibrium states are visited
successively. The states can be characterized by the latitudinal
position of their northern ice edges in winter and summer,
respectively (Fig.\ \ref{fig:hysteresis}). First, simulations start at
low stellar irradiance $\tilde{S}=0.7$ in the fully ice covered
cryoplanet state.  Once the northern edge of the summer sea ice cover
moves southward the polar ocean becomes seasonally ice free, while
winter sea ice still extends from the equator to the pole. This is the
near cryoplanet state. Upon further increasing $\tilde{S}$, also the
northern edge of the winter sea ice cover begins to move equatorward
and the polar oceans now remain open throughout the year. This is the
Uncapped Cryoplanet that is stable for a considerable range of stellar
irradiance. The gradual equatorward movement of both northern ice edge
latitudes when irradiance becomes stronger eventually produces an ice
free ocean, the aquaplanet state. Note that in
Fig.\ \ref{fig:hysteresis}c the hysteresis branch for the winter sea
ice edge transition from 90\textdegree\ to 0\textdegree\ is not
vertical which would suggest a very narrow range of an uncapped
cryoplanet. However, such states are only transient, and the finite
slope of the hysteresis branch is caused by the rate of change of
$\tilde{S}$ being faster than the approach to the steady state of the
Aquaplanet.

Second, when $\tilde{S}$ is reduced again to below 0.9, a tipping
point for both the winter and summer ice edge is reached
simultaneously. While the northern edge position of the summer sea ice
cover arrests at about 50\textdegree\ latitude and then moves
poleward, the winter ice edge position jumps directly to
90\textdegree, i.e., the planet makes a transition from a perennially
ice free to a completely ice covered ocean during winter.  Therefore,
no uncapped cryoplanet state is found when simulations start from an
Aquaplanet. An important result is that irrespective of $K_h$ the
uncapped cryoplanet state is not reached when starting from an
Aquaplanet. This suggests that the particulars of the ocean model are
not the main reason why other studies did not find stable states with
an equatorial ice belt at high obliquity
\citep{Chandler2000,Williams2003,Ferreira2014,Jenkins2000,Jenkins2001,Jenkins2003},
but rather the albedo effect afforded by the initial conditions.

\section{Conclusion}\label{sec:Conclusion}

We have shown using a dynamical atmosphere-ocean climate model that
the state with an equatorial ice belt, the Uncapped Cryoplanet, is a
stable state at high obliquity for a range of stellar irradiance. This
confirms our earlier findings that used a simpler configuration
consisting of an atmospheric model coupled to a slab ocean
\citep{Kilic2017b}. Our results apply to both the slab and the
dynamical ocean versions. Furthermore, our simulations using the
simpler configuration as an emulator model provide insight into the
stability of the uncapped cryoplanet state and its dependence on
initial conditions. With increasing importance of the ocean heat
transport the equatorial ice cover shrinks in latitudinal extent until
it becomes unsustainable. This marks a critical point where a
transition to the aquaplanet state occurs. Transient simulations in
which the stellar irradiance is slowly increased or decreased,
demonstrate that the Uncapped Cryoplanet only is reached from
more glaciated conditions, e.g.\ the cryoplanet or the near cryoplanet
states, but not from less glaciated states. This result does not
depend on the presence or absence of a meridional ocean heat flux.

This is reminiscent of robust hysteresis behavior and multiple
equilibria in the non-linear atmosphere-ice-ocean system, as
demonstrated earlier \citep{Lucarini2013,Kilic2017b,Rose2017}. We have
shown that the shape of the hysteresis depends on model parameters. By
inference, one would expect that the position and shape of a
hysteresis loop is also model dependent. \citet{Ferreira2014}
concluded that an equatorial ice belt is unlikely to exist. The
difference to our finding could be due to two reasons. First, their
model consisted of an ocean component based on primitive equations
which is dynamically more realistic than the LSG formulation
which neglects the non-linear momentum advection.  This would likely
result in a different hysteresis structure. Second and more
importantly, they initialized their simulations only from an
aquaplanet state. Hence, the Uncapped Cryoplanet may also be realized
in their model if their simulations had started from a more glaciated
state. Further exploration of the solution space of that model is
therefore warranted. More generally, we suggest that exoplanet
research would highly benefit from coordinated model intercomparison
efforts, as have now become the standard in the Earth System Model
community \citep{Eyring2016}.

We emphasize that our simulations are idealized and only represent a
first approach to the investigation of stable climate regimes under
high obliquity. This is justifiable in the absence of more detailed
information on exoplanet surface conditions and configurations. For
instance, during an exoplanet's evolution tectonic processes could
form ocean basins and land masses that would fundamentally alter the
ocean circulation and the associated meridional heat flux. While land
masses would provide a solid ground for growing ice masses and enhance
the albedo feedbacks, they would also be more susceptible to heating
due to their smaller heat capacity. Therefore, it is currently not
possible to assess the consequence of different land-ocean
configurations on the distribution of stable climate states on a
high-obliquity exoplanet. Further observational information and
simulations are required to explore such questions.

In conclusion, our simulations demonstrate that the irradiation
history of an exoplanet is the primary determinant of whether an
equatorial ice belt would occur as a climate state during its
evolution. Once the planet has reached an ice free state, it seems
unlikely that it would reach an uncapped cryoplanet again at some
later stage, unless another planet-scale glaciation has occurred. This
finding is relevant for understanding habitable conditions over the
entire climatic evolution of an exoplanet.

\acknowledgments

We acknowledge continuous model support by E. Kirk and constructive
comments by an anonymous reviewer. Simulations were performed on
UBELIX, the HPC cluster at University of Bern, and at the Swiss
National Supercomputing Center (CSCS). FL acknowledges support from
the Cluster of Excellence for Integrated Climate Science (CLISAP). CCR
and TFS received support from the Swiss National Science Foundation.

%\bibliography{biblio2}

\end{document}